\documentclass[12pt]{article}
  \usepackage{amsthm,amsfonts}
  \usepackage{amsmath}
\newcommand{\bea}   {\begin{eqnarray}}
\newcommand{\eea}   {\end{eqnarray}}
\begin{document}
\renewcommand{\thefootnote}{\fnsymbol{footnote}}

\thispagestyle{empty}

\title{Twist Deformations of the\\ Supersymmetric Quantum Mechanics}
\author{P. G. Castro\thanks{{\em e-mail: pgcastro@cbpf.br}},~
B. Chakraborty\thanks{{\em e-mail: biswajit@bose.res.in}},~
Z. Kuznetsova\thanks{{\em e-mail: zhanna.kuznetsova@ufabc.edu.br}}
 ~and F. Toppan\thanks{{\em e-mail: toppan@cbpf.br}}
\\ \\
{\it $~^{\ast\dagger\S}$ CBPF, Rua Dr.} {\it Xavier Sigaud 150,} \\{\it cep 22290-180, Rio de Janeiro (RJ), Brazil.}\\
{\it $~^\dagger$ S. N. Bose National Center for Basic Sciences,}\\ {\it JD Block, Sector III, Salt-Lake, Kolkata-700098, India.}
\\
{\it $~^{\ddagger}$ UFABC, Rua Catequese 242, Bairro Jardim,} \\{\it  cep 09090-400, Santo Andr\'e (SP), Brazil.}}
\maketitle
\begin{abstract}
The ${\cal N}$-extended Supersymmetric Quantum Mechanics is deformed via an abelian twist which preserves the super-Hopf algebra structure of its Universal Enveloping Superalgebra. Two constructions are possible.
For even ${\cal N}$ one can identify the $1D$ ${\cal N}$-extended superalgebra with the fermionic Heisenberg algebra. Alternatively, supersymmetry generators can be realized as operators belonging to the Universal Enveloping Superalgebra of one bosonic and several fermionic oscillators. \par
The deformed system is described in terms of twisted operators satisfying twist-deformed (anti)commutators.\par
The main differences between an abelian twist defined in terms of fermionic operators and an abelian twist defined in terms of bosonic operators are discussed. \par
\end{abstract}
\vfill

\rightline{CBPF-NF-012/09}

\newpage
\section{Introduction}
In this paper we investigate the abelian twist-deformation of the fermionic Heisenberg algebra,
introduced in \cite{cct}, in application to the deformation of the Supersymmetric Quantum Mechanics.
We remark that the abelian twist deformation of the fermionic Heisenberg algebra gives rise to the Cliffordization of the Grassmann variables, recovering, in a more general setting, the results of
\cite{hh}. The connection of the fermionic Heisenberg algebra with the graded algebra underlying the
Supersymmetric Quantum Mechanics is twofold. Indeed, the ${\cal N}$-extended, one-dimensional superalgebra is isomorphic to the fermionic Heisenberg algebra for an even number ${\cal N}$ of (odd) supercharges. On the other hand, the same superalgebra can be realized, essentially, in terms of operators belonging to the Universal Enveloping Algebra of one bosonic Heisenberg algebra and several copies of the fermionic Heisenberg algebras. In the following we point out the differences of the two
schemes. Within the second scheme we can recover, on the module, a lower dimensional version of the Cliffordization
of  \cite{ks} and \cite{ihl}.\par
We further point out that in all previous studies in the literature graded undeformed brackets or graded, Moyal-type, brackets were used along with the deformed coproducts. In our approach, on the other hand, motivated by the considerations in \cite{ckt} (restricted, in that paper, to bosonic algebras), we use twist-deformed brackets, as in \cite{cct}. The twist-deformed brackets, which implement twist-deformed adjoint action, are directly associated with the twist-deformed coproduct. \par
The abelian character of the (fermionic) twist deformation implies that the fermionic supercharges are deformed. On the other hand, the bosonic charges remain undeformed. Nevertheless, even in the bosonic sector, the theory gets modified since the coproduct applied to bosonic operators gets deformed and can therefore affect multi-particle bosonic systems. In order to stress the difference between a ``fermionic'' abelian twist and a ``bosonic'' abelian twist (in the first case the exponential defining the twist is given by a tensor product of fermionic generators while, in the second case, it is a tensor product of bosonic generators), we apply the abelian twist deformation
to (an enlarged version of) the bosonic Heisenberg algebra for a particularly simple example. We show that in this case even the bosonic operators get deformed, not just their coproducts. In particular, an operator which gets deformed is the Hamiltonian of the harmonic oscillator. \par
Returning to Supersymmetric Quantum Mechanics, in the second framework (generators realized as operators in a Universal Enveloping Superalgebra), besides the supercharges, the fermionic derivative operators can be constructed (and deformed). The fact that their coproduct is deformed can be interpreted as a breaking of the (graded) Leibniz rule, in accordance with several results obtained in the literature, by using a variety of different methods \cite{zup,krt,drw}.\par
Non-anticommutative supersymmetric theories have been investigated by assuming, either as a
mathematical possibility or in the string context, the spinorial coordinates
to be non-anticommutative (see \cite{svn, fl, flm, kpt, bgn, ov1, ov2, sei}).
The particularly influential \cite{sei} paper introduced non-anticommutative supersymmetry in a $4$-dimensional
Euclidean superspace. It provided motivations for studying lower-dimensional non-anticommutative supersymmetric models \cite{ck, chandra, hkks1, hkks2, agvm, fgnps}. In one time-dimension, in particular, (i.e. non-relativistic Supersymmetric Quantum Mechanics) non-anticommutative deformations were studied in \cite{as,is,krt}.\par
It comes as no surprise that the majority of the works on non-anticommutative supersymmetry found inspiration from the related works of non-commutative deformations of the ordinary bosonic theories (see \cite{bal} for a review), whose recent upsurge is essentially due to the seminal works
of \cite {dfr} and \cite {sw}. The notion of the Drinfeld twist (of the
Universal Enveloping Algebra of the Poincar\'e algebra) was first applied in \cite{cha} to
restore the Lorentz symmetry which would be otherwise spoiled in a relativistic non-commutative theory.
In the supersymmetric context, the Drinfeld twist has been investigated in the already recalled papers \cite{ihl,ks,zup,drw}, while a Jordanian twist for the $osp(1|2)$ Superconformal Quantum Mechanics was also considered in \cite{blt}.\par
The scheme of this paper is as follows. In the next Section we introduce the abelian twist of the fermionic Heisenberg algebra, presenting all the relevant formulas (twist-brackets and so on).  In Section {\bf 3} we point out the isomorphism between fermionic Heisenberg algebra and the one-dimensional ${\cal N}$-Extended superalgebra for even values of ${\cal N}$. The twist-deformation of the Supersymmetric Quantum Mechanics in terms of its superspace representation is investigated in Section {\bf 4}. We introduce the relevant Universal Enveloping Algebras for the construction. The specializations to ${\cal N}=2$ and ${\cal N}=4$ are detailed. Non-upper triangular supersymmetry generators obtained from twist are given.  The abelian twist in the bosonic case is presented in Section {\bf 5}.

\section{The twist deformation of the fermionic Heisenberg algebra}

The Grassmann algebra is generated by $\mathcal{N}$ anticommuting coordinates $\theta_\alpha$.  These coordinates, together with their fermionic, Berezin, derivatives $\partial_\alpha$, define a Lie superalgebra with $2\mathcal{N}$ odd generators (the Grassmann coordinates and their derivatives) and a single even generator,
a central charge $z$. The (anti)-commutation relations are given by
\begin{eqnarray}\label{grassmann}
  \left\{\theta_\alpha,\theta_\beta\right\}&=&\left\{\partial_\alpha,\partial_\beta\right\}=0, \nonumber\\
  \left\{\partial_\alpha,\theta_\beta\right\}&=&\delta_{\alpha\beta}z,\nonumber\\
  \left[z,\partial_\alpha\right]&=&\left[z,\theta_\alpha\right]=0.
\end{eqnarray}

The central charge has to be introduced in order to form a Lie superalgebra.

The (\ref{grassmann}) algebra is the fermionic Heisenberg algebra and will be denoted, see \cite{cct}, as $h_F(\mathcal{N})$. Its Universal Enveloping Algebra $\mathcal{U}(h_F(\mathcal{N}))$ has the structure of a Hopf superalgebra. The central charge $z$ is treated at par
with the other generators $\theta_\alpha$ and $\partial_\alpha$\footnote{This is analogous to considering the mass parameter of the Galileo algebra as a Lie generator, so that one can use non-projective representation \cite{wein, sor}. The same prescription was used in \cite{cct} to deform the Heisenberg algebra.},
so its coproduct is $\Delta(z)=z\otimes
\mathbf{1}+\mathbf{1}\otimes z$ and the antipode is $S(z)=-z$. In the undeformed case the coproduct of $\theta_\alpha$ and
$\partial_\alpha$ reads, respectively,
$\Delta(\theta_\alpha)=\theta_\alpha\otimes
\mathbf{1}+\mathbf{1}\otimes \theta_\alpha$ and $\Delta(\partial_\alpha)=\partial_\alpha\otimes
\mathbf{1}+\mathbf{1}\otimes \partial_\alpha$. The corresponding antipodes are given as $S(\theta_\alpha)=-\theta_\alpha$ and $S(\partial_\alpha)=-\partial_\alpha$

A mass-dimension can be assigned to the generators. We have
$[\theta_\alpha]=-\frac{1}{2}$, $[\partial_\alpha]=\frac{1}{2}$, $[z]=0$.

An abelian twist ${\mathcal F}$ (for a review, see \cite{aschieri} and references therein) given by

\begin{eqnarray}\label{abtwist}
    \mathcal{F}&:= &f^\alpha\otimes f_\alpha=\exp\left(C_{\alpha\beta}\partial_\alpha\otimes\partial_\beta \right), \nonumber\\
    {\mathcal{F}}^{-1} &:=& {\overline f}^\alpha\otimes{\overline f}_\alpha= \exp\left(-C_{\alpha\beta}\partial_\alpha\otimes\partial_\beta \right)
\end{eqnarray}
 can be introduced in terms of the diagonal matrix
\bea
C_{\alpha\beta}&=&\frac{1}{M}\eta_{\alpha\beta},
\eea
where $M$ is a mass-parameter and $\eta_{\alpha\beta}$ is a non-dimensional matrix which admits $p$ positive diagonal elements $+1$, $q$  negative diagonal elements
$-1$ and $r$ zero elements ($p+q+r=\mathcal{N}$).

(the Sweedler notation has been used).\par
Following the notations and conventions of \cite{cct}, the only deformed coproducts correspond to $\theta_\alpha$
(the others remain undeformed):
\begin{equation}
    \Delta^\mathcal{F}(\theta_\alpha)={\cal F}\Delta(\theta_\alpha){\cal F}^{-1}=\Delta(\theta_\alpha)+C_{\alpha\beta}(\partial_\beta\otimes z -
    z\otimes\partial_\beta),
\end{equation}
where $\Delta(\theta_\alpha)=\theta_\alpha\otimes\mathbf{1}+\mathbf{1}\otimes\theta_\alpha$ is the undeformed coproduct introduced earlier.

Since
\begin{equation}
    \chi=f^\alpha
    S(f_\alpha)=\exp\left(-C_{\alpha\beta}\partial_\alpha\partial_\beta\right)=\mathbf{1},
\end{equation}
the antipode is also undeformed.

Among the generators, only the $\theta_\alpha$'s get deformed. We have

\begin{equation} \label{defteta}
    \theta_\alpha^\mathcal{F}:= {\overline f}^\beta(\theta_\alpha){\overline f}_\beta=\theta_\alpha+C_{\alpha\beta}\partial_\beta
    z.
\end{equation}
The deformed generators differ from the original ones in terms of a shift (the fermionic counterpart of the Bopp shift, see also \cite{bls}).

The universal $\mathcal{R}$-matrix is simply $\mathcal{F}^{-2}$,
so that
\begin{equation}
    \Delta^\mathcal{F}(\theta_\alpha^\mathcal{F})=
    \theta_\alpha^\mathcal{F}\otimes\mathbf{1}+\mathbf{1}\otimes\theta_\alpha^\mathcal{F}+2C_{\alpha\beta}\partial_\beta\otimes
    z.
\end{equation}

The antipodes are
\begin{equation}
    S(\theta_\alpha^\mathcal{F})=-\theta_\alpha+C_{\alpha\beta}z\partial_\beta
    =-\theta_\alpha^\mathcal{F}+2C_{\alpha\beta}z\partial_\beta.
\end{equation}

The deformed brackets $$\left[u^\mathcal{F},v^\mathcal{F}\right\}_\mathcal{F}=\sum_k(u^\mathcal{F})^k_1v^\mathcal{F}(-1)^{|v^\mathcal{F}||(u^\mathcal{F})^k_2|}S(u^\mathcal{F})^k_2$$ of the deformed generators coincide with the original (\ref{grassmann}) algebra:

\begin{eqnarray}
  \left\{\theta_\alpha^\mathcal{F},\partial_\beta^\mathcal{F}\right\}_\mathcal{F}
  &=&\delta_{\alpha\beta}z^\mathcal{F}, \nonumber\\
  \left\{\theta_\alpha^\mathcal{F},\theta_\beta^\mathcal{F}\right\}_\mathcal{F}&=&0,\nonumber \\
  \left\{\partial_\alpha^\mathcal{F},\partial_\beta^\mathcal{F}\right\}_\mathcal{F}&=&0,\nonumber \\
  \left[\partial_\alpha^\mathcal{F},z^\mathcal{F}\right]_\mathcal{F}
  &=&\left[\theta_\alpha^\mathcal{F},z^\mathcal{F}\right]_\mathcal{F}=0.
\end{eqnarray}

The ordinary brackets of the deformed quantities yield a nonlinear algebra
\begin{eqnarray}
  \left\{\theta_\alpha^\mathcal{F},\partial_\beta^\mathcal{F}\right\}&=&\delta_{\alpha\beta}z,\nonumber \\
  \left\{\theta_\alpha^\mathcal{F},\theta_\beta^\mathcal{F}\right\}&=&2C_{\alpha\beta}z^2,\nonumber \\
  \left\{\partial_\alpha^\mathcal{F},\partial_\beta^\mathcal{F}\right\}&=&0,\nonumber \\
  \left[z^\mathcal{F},\theta_\alpha^\mathcal{F}\right]
  &=&\left[z^\mathcal{F},\partial_\alpha^\mathcal{F}\right]=0.
  \label{d4}
\end{eqnarray}

The multiplication $m$ on the module \cite{cct} acts as follows
\bea
&m(\theta_\alpha \otimes \theta_\beta)= \theta_\alpha\cdot\theta_\beta,\quad m(\theta_\alpha \otimes z)= \theta_\alpha\cdot z,
\quad m(z\otimes z)= z^2.&
\eea
The deformed multiplication between elements belonging to the module (using the Sweedler notation) is given by
\begin{eqnarray}
    a\star b\equiv m^\mathcal{F}(a\otimes b)=(m\circ\mathcal{F}^{-1})(a\otimes
    b)=\sum_\alpha(-1)^{|\bar{f}_\alpha||a|}\bar{f}^\alpha(a)\bar{f}_\alpha(b).
\end{eqnarray}

Defining $\left[a,b\right\}_\star\equiv a\star b
+(-1)^{|a||b|} b\star a$, we have
\begin{eqnarray}
  \left\{ \theta_\alpha , \theta_\beta\right\}_\star&=&2C_{\alpha\beta}z^2, \nonumber\\
  \left[z,\theta_\alpha\right]_\star&=&0.
\end{eqnarray}
The above fermionic Moyal-brackets, together with the relations
\bea
  \left\{\partial_\alpha , \theta_\beta\right\}_\star&=&\delta_{\alpha\beta}z,\nonumber \\
  \left\{\partial_\alpha , \partial_\beta\right\}_\star&=&0,\nonumber\\
  \left[z,\partial_\alpha\right]_\star&=&0,
\eea
makes the $\star$-brackets
isomorphic to the ordinary brackets of the deformed quantities.

\section{The $1D$ $\cal{N}$-Extended Superalgebra}

In this section, we are going to basically establish the isomorphism between the fermionic Heisenberg algebra introduced in the previous section and the one-dimensional $\mathcal{N}$-extended superalgebra for even values of $\mathcal{N}$. To this end, consider the supersymmetry algebra with odd generators ${\widehat Q}_I$ ($I,J=1,\ldots, {\cal N}$) and an even generator $H$, given by
\begin{eqnarray}\label{susy}
  \{\widehat{Q}_I, \widehat{Q}_J\} &=& \delta_{IJ}H,\nonumber\\
    \left[H,\widehat{Q}_I\right] &=& 0.
\end{eqnarray}
For ${\cal N}$ even, we can split the odd sector into a chiral and an antichiral set:
\begin{eqnarray}
  Q_i &=& \widehat{Q}_i+i\widehat{Q}_{i+\frac{{\cal N}}{2}}, \nonumber\\
  \overline{Q}_i &=& \widehat{Q}_i-i\widehat{Q}_{i+\frac{{\cal N}}{2}},
  \end{eqnarray}
with $i=1,\ldots,\frac{{\cal N}}{2}$. \par
The algebra can be reexpressed as
\begin{eqnarray}\label{susy2}
\left\{Q_i,\overline{Q}_j\right\}&=&2\delta_{ij}H,\nonumber\\
  \left\{Q_i,Q_j\right\}&=&\left\{\overline{Q}_i,\overline{Q}_j\right\}=0,\nonumber\\
  \left[H,Q_i\right]&=&\left[H,\overline{Q}_i\right]=0.
\end{eqnarray}

It is isomorphic to (\ref{grassmann}) if we identify $Q_i$ with $\theta_\alpha$, $\overline{Q}_i$ with $\partial_\alpha$ and $2H$ with $z$.

We shall deform the algebra (\ref{susy2}) by means of the Abelian twist
\begin{equation}
    \mathcal{F}=\exp\left(\frac{C_{ij}}{2}\overline{Q}_i\otimes\overline{Q}_j \right),
\end{equation}
with $C_{ij}=\frac{\eta_{ij}}{M}$, where $\eta_{ij}$ is a non-dimensional diagonal matrix admitting $p$ positive +1 , $q$ negative -1 and $r$ zero entries ($p+q+r={\cal N}$). \par
This deformation coincides with (\ref{abtwist}). In particular the deformed coproduct of ${Q_i}$ reads
\begin{equation}
    \Delta^\mathcal{F}({Q}_i)=\Delta({Q}_i)+C_{ij}(\overline{Q}_j\otimes H-H\otimes \overline{Q}_j).
\end{equation}
The antipode is undeformed due to
\begin{equation}
    \chi=f^\alpha
    S(f_\alpha)=\exp\left(-\frac{C_{ij}}{2}\overline{Q}_i \overline{Q}_j\right)=\mathbf{1}.
\end{equation}
The only deformed generators are the $Q_i$:
\begin{equation}
    {Q}_i^\mathcal{F}=Q_i+C_{ij}\overline{Q}_j
    H.
\end{equation}

The universal $\mathcal{R}$-matrix is  $\mathcal{F}^{-2}$, so
\begin{equation}
    \Delta^\mathcal{F}(Q_i^\mathcal{F})=Q_i^\mathcal{F}\otimes\mathbf{1}+\mathbf{1}\otimes Q_i^\mathcal{F}+2C_{ij}{\overline{Q}}_j\otimes
    H.
\end{equation}

The antipodes are
\begin{equation}
    S(Q_i^\mathcal{F})=-Q_i+C_{ij}\overline{Q}_j
    H=-Q_i^\mathcal{F}+2C_{ij}\overline{Q}_j
    H.
\end{equation}
Now the deformed brackets are:

\begin{eqnarray}
  \left\{\overline{Q}_i^\mathcal{F},Q_j^\mathcal{F}\right\}_\mathcal{F}&=&\delta_{ij}H^\mathcal{F}=\delta_{ij}H,\nonumber\\
  \left\{\overline{Q}_i^\mathcal{F},\overline{Q}_j^\mathcal{F}\right\}_\mathcal{F}&=&0,\nonumber \\
  \left\{Q_i^\mathcal{F},Q_j^\mathcal{F}\right\}_\mathcal{F}&=&0,\nonumber \\
  \left[Q_i^\mathcal{F},H^\mathcal{F}\right]_\mathcal{F}&=&\left[\overline{Q}_i^\mathcal{F},H^\mathcal{F}\right]_\mathcal{F}=0.
\end{eqnarray}
The ordinary brackets of the deformed quantities are
\begin{eqnarray}\label{mixsusy}
  \left\{\overline{Q}_i^\mathcal{F},Q_j^\mathcal{F}\right\}&=&\delta_{ij}H^\mathcal{F},\nonumber\\
  \left\{\overline{Q}_i^\mathcal{F},\overline{Q}_j^\mathcal{F}\right\}&=&0,\nonumber \\
  \left\{Q_i^\mathcal{F},Q_j^\mathcal{F}\right\}&=&2C_{ij}({H^{\mathcal{F}}})^2,\nonumber \\
  \left[H^\mathcal{F},\overline{Q}_i^\mathcal{F}\right]&=&\left[H^\mathcal{F},Q_i^\mathcal{F}\right]=0.
\end{eqnarray}

\section{The superspace representation}

To make the connection with the superspace we need to introduce the Grassmann variables $\theta_I$ and their derivatives $\partial_{\theta_I}$
which (along with the central extension $z$) satisfy the $h_F(N)$ algebra, as well as the bosonic parameter $t$ and
its derivative $\partial_t$ which (along with the central extension ${\hbar}$) satifies the bosonic Heisenberg
algebra $h_B(1)$, obtaining, in principle, the $h_B(1)\oplus h_F(N)$
algebra. We can now identify the two central extensions ($z=\hbar$), thereby obtaining an algebra which we shall call $h(1,N)$. Throughout this section we will
be working with its universal enveloping algebra, ${\cal U}(h(1,N))$.
An explicit realization of the ${\cal N}$-extended supersymmetry algebra (\ref{susy})
is obtained in terms of composite operators belonging to ${\cal U}(h(1,\mathcal{N}))$
(${\cal N}\equiv N$)\footnote{We allow Laurent-expansion of the central element $\hbar$. Concerning the undeformed coproduct of $\widehat{Q}_I$ introduced below, it is assumed to coincide with the undeformed coproduct of a fermionic primitive element, i.e., $\Delta({\widehat Q}_I)=\widehat{Q}_I\otimes \mathbf{1}+\mathbf{1}\otimes\widehat{Q}_I$, so that one never encounters ambiguous expressions like $\Delta(\frac{1}{\hbar})$ in (\ref{q}).}. Explicitly,

\begin{eqnarray}\label{q}
    \widehat{Q}_I &=& \partial_{\theta_I}+\frac{i} {\hbar}{\theta_I\partial_t},\nonumber\\
    H &=& i\partial_t,
\end{eqnarray}
with $I=1,\ldots,{\cal N}$.

Since we are working with ${\cal U}(h(1,\mathcal{N}))$, it is now possible to apply the twist (\ref{abtwist})

\begin{equation}
\mathcal{F}=\exp(C_{IJ}\partial_{\theta_I}\otimes\partial_{\theta_J})
\end{equation}
to the supersymmetry generators.

We obtain that the deformed coproduct of  $\widehat{Q}_I$ is
\begin{equation}
    \Delta^\mathcal{F}(\widehat{Q}_I)=\Delta(\widehat{Q}_I)+C_{IJ} (\partial_{\theta_J}\otimes H-H\otimes\partial_{\theta_J}).
    \end{equation}

The deformed generators are
\begin{equation}
    \widehat{Q}_I^\mathcal{F}=\widehat{Q}_I+C_{IJ}H\partial_{\theta_J},
\end{equation}
the deformed coproduct of the deformed generators being
\begin{equation}
    \Delta^\mathcal{F}(\widehat{Q}_I^\mathcal{F})=\widehat{Q}_I^\mathcal{F}\otimes \mathbf{1}+\mathbf{1}\otimes\widehat{Q}_I^\mathcal{F}+2C_{IJ} (\partial_{\theta_J}\otimes H).
    \end{equation}

    The ordinary brackets of the deformed generators read
        \begin{equation}\label{nlinsusy2}
        \{\widehat{Q}_I^\mathcal{F},\widehat{Q}_J^\mathcal{F}\}=\delta_{IJ}H+2C_{IJ}H^2,
    \end{equation}
whereas the deformed brackets read just
   \begin{equation}
        \{\widehat{Q}_I^\mathcal{F},\widehat{Q}_J^\mathcal{F}\}_\mathcal{F}=\delta_{IJ}H.
    \end{equation}

In a different context, non-linear supersymmetry such as encountered in (\ref{nlinsusy2}) or in the r.h.s. of (\ref{mixsusy})
was discussed in \cite{ply}.

\subsection{The ${\cal N}=2$ case}

Let us now consider the ${\cal U}(h(1,2))$-algebra realization of the ${\cal N}=2$ supersymmetry generators
\begin{eqnarray}
  \widehat{Q}_1 &=& \partial_{\theta_1}+\frac{i}{\hbar}\theta_1\partial_t,\nonumber  \\
  \widehat{Q}_2 &=& \partial_{\theta_2}+\frac{i}{\hbar}\theta_2\partial_t.
  \end{eqnarray}
These can be further augmented by the fermionic covariant derivatives
  \begin{eqnarray}
  D_1 &=& \partial_{\theta_1}-\frac{i}{\hbar}\theta_1\partial_t, \nonumber \\
  D_2 &=& \partial_{\theta_2}-\frac{i}{\hbar}\theta_2\partial_t.
\end{eqnarray}

They satisfy the so-called ${\cal N}=(2,2)$ pseudo-supersymmetry algebra

\begin{eqnarray}
  \{\widehat{Q}_I, \widehat{Q}_J\} &=& \delta_{IJ}H,\nonumber \\
  \{D_I,D_J\} &=& -\delta_{IJ}H,\nonumber \\
  \{D_I,\widehat{Q}_J\} &=& 0,\nonumber \\
  \left[H, \widehat{Q}_I\right] &=& \left[H, D_I\right]=0.
\end{eqnarray}

To deform it we can now apply any twist $\mathcal{F}\in\mathcal{U}(h(1,2))\otimes\mathcal{U}(h(1,2))$ which is invertible and satisfies the cocycle condition.

An acceptable abelian twist is
\begin{equation}
    \mathcal{F}=\exp\left(\frac{\epsilon}{M}\overline{Q}\otimes\overline{Q}+\frac{\eta}{M}\overline{D}\otimes\overline{D} \right),
\end{equation}
where
\begin{eqnarray}
  \overline{Q} &=& \widehat{Q}_1-i\widehat{Q}_2, \nonumber\\
  \overline{D} &=& D_1-iD_2
\end{eqnarray}
and $\epsilon$, $\eta$ are numbers which can be normalized, without loss of generality, to be $+1$, $-1$ or $ 0$
(the abelian twist described in Section {\bf 3} is recovered for $\eta=0$, $C_{11}=\frac{\epsilon}{M}$). \\
This twist trivially satisfies the cocycle condition since $\{\overline{Q},\overline{Q}\}=\{\overline{Q},\overline{D}\}=\{\overline{D},\overline{D}\}=0$.

The deformations of the generators are given by

\begin{eqnarray}
  \widehat{Q}_1^\mathcal{F} &=& \widehat{Q}_1+\frac{\epsilon}{M}(\widehat{Q}_1-i\widehat{Q}_2)H, \nonumber\\
  \widehat{Q}_2^\mathcal{F} &=& \widehat{Q}_2-\frac{i\epsilon}{M}(\widehat{Q}_1-i\widehat{Q}_2)H,\nonumber \\
  D_1^\mathcal{F} &=& D_1+\frac{\eta}{M}(D_1-iD_2)H,\nonumber  \\
  D_2^\mathcal{F} &=& D_2-\frac{i\eta}{M}(D_1-iD_2)H.
\end{eqnarray}
together with $H^{\mathcal F}=H$.
The deformed coproducts are

\begin{eqnarray}
  \Delta^\mathcal{F}(\widehat{Q}_1) &=& \Delta(\widehat{Q}_1)+\frac{\epsilon}{M}(\widehat{Q}_1\otimes H-H\otimes \widehat{Q}_1)-\frac{i\epsilon}{M}(\widehat{Q}_2\otimes H-H\otimes \widehat{Q}_2),\nonumber \\
   \Delta^\mathcal{F}(\widehat{Q}_2) &=& \Delta(\widehat{Q}_2)-\frac{i\epsilon}{M}(\widehat{Q}_1\otimes H-H\otimes \widehat{Q}_1)-\frac{\epsilon}{M}(\widehat{Q}_2\otimes H-H\otimes \widehat{Q}_2),\nonumber\\
   \Delta^\mathcal{F}(D_1) &=& \Delta(D_1)-\frac{\eta}{M}(D_1\otimes H-H\otimes D_1)+\frac{i\eta}{M}(D_2\otimes H-H\otimes D_2),\nonumber\\
 \Delta^\mathcal{F}(D_2) &=& \Delta(D_2)+\frac{i\eta}{M}(D_1\otimes H-H\otimes D_1)+\frac{\eta}{M}(D_2\otimes H-H\otimes D_2).
\end{eqnarray}

The antipode does not get deformed since

\begin{equation}
    \chi=f^\alpha S(f_\alpha)=\exp\left(-\frac{\epsilon}{M} \overline{Q}^2-\frac{\eta}{M} \overline{D}^2 \right)=\mathbf{1},
\end{equation}

so that the antipodes are
\begin{eqnarray}
  S(\widehat{Q}_1^\mathcal{F}) &=& -\widehat{Q}_1^\mathcal{F}+\frac{2\epsilon}{M}(\widehat{Q}_1-i\widehat{Q}_2),\nonumber \\
  S(\widehat{Q}_2^\mathcal{F}) &=& -\widehat{Q}_2^\mathcal{F}-\frac{2i\epsilon}{M}(\widehat{Q}_1-i\widehat{Q}_2),\nonumber  \\
  S(D_1^\mathcal{F}) &=& -D_1^\mathcal{F}-\frac{2\eta}{M}(D_1-iD_2),\nonumber  \\
  S(D_2^\mathcal{F}) &=& -D_2^\mathcal{F}+\frac{2i\eta}{M}(D_1-iD_2).
\end{eqnarray}

The universal $\mathcal{R}$-matrix is

\begin{equation}
    \mathcal{R}=\mathcal{F}^{-2}=\exp\left[-2\left(\frac{\epsilon}{M}\overline{Q}\otimes\overline{Q}+\frac{\eta}{M}\overline{D}\otimes\overline{D} \right) \right].
\end{equation}
With those, we are able to work out the deformed coproducts of the deformed quantities. They are
\begin{eqnarray}
  \Delta^\mathcal{F}(\widehat{Q}_1^\mathcal{F}) &=& \widehat{Q}_1^\mathcal{F}\otimes\mathbf{1}+\mathbf{1}\otimes \widehat{Q}_1^\mathcal{F}+\frac{2\epsilon}{M}(\widehat{Q}_1-i\widehat{Q}_2)\otimes H,\nonumber \\
  \Delta^\mathcal{F}(\widehat{Q}_2^\mathcal{F}) &=& \widehat{Q}_2^\mathcal{F}\otimes\mathbf{1}+\mathbf{1}\otimes \widehat{Q}_2^\mathcal{F}-\frac{2i\epsilon}{M}(\widehat{Q}_1-i\widehat{Q}_2)\otimes H,\nonumber  \\
  \Delta^\mathcal{F}(D_1^\mathcal{F}) &=& D_1^\mathcal{F}\otimes\mathbf{1}+\mathbf{1}\otimes D_1^\mathcal{F}-\frac{2\eta}{M}(D_1-iD_2)\otimes H ,\nonumber \\
  \Delta^\mathcal{F}(D_2^\mathcal{F}) &=& D_2^\mathcal{F}\otimes\mathbf{1}+\mathbf{1}\otimes D_2^\mathcal{F}+\frac{2i\eta}{M}(D_1-iD_2)\otimes H .
\end{eqnarray}

These yield, as expected, the deformed brackets of the deformed quantities:

\begin{eqnarray}
  \{\widehat{Q}_I^\mathcal{F}, \widehat{Q}_J^\mathcal{F}\}_\mathcal{F} &=& \delta_{IJ}H^\mathcal{F},\nonumber \\
  \{D_I^\mathcal{F},D_J^\mathcal{F}\}_\mathcal{F} &=& -\delta_{IJ}H^\mathcal{F},\nonumber \\
  \{D_I^\mathcal{F},\widehat{Q}_J^\mathcal{F}\}_\mathcal{F} &=& 0.
\end{eqnarray}

We now want to study the deformed multiplication on a module consisting of the space of functions of Grassmann variables $\theta_1, \theta_2$. The ordinary multiplication $m$ acts as the usual Grassmann product, that is

\begin{equation}
    m(\theta_I\otimes\theta_J)=\theta_I\cdot\theta_J.
\end{equation}

The action of $\widehat{Q}_I$ and $D_I$ is

\begin{eqnarray}
  &\{\widehat{Q}_I,\theta_J\}=
  \{D_I,\theta_J\}=\delta_{IJ}.
\end{eqnarray}

Now we define the star product to be

\begin{equation}
    \theta_I\star\theta_J=m^\mathcal{F}(\theta_I\otimes\theta_J)=(m\circ\mathcal{F}^{-1})(\theta_I\otimes\theta_J)
\end{equation}
and proceed to calculate explicitly
\begin{eqnarray}
  \theta_1\star\theta_1&=& -\frac{\epsilon}{M} - \frac{\eta}{M},\nonumber\\
   \theta_2\star\theta_2&=& \frac{\epsilon}{M} + \frac{\eta}{M},\nonumber\\
    \theta_1\star\theta_2&=& -\frac{i\epsilon}{M} - \frac{i\eta}{M}+\theta_1\theta_2,
\end{eqnarray}

so that the star-anticommutators are
\begin{eqnarray}
  \{\theta_1,\theta_1\}_\star&=& -2\left(\frac{\epsilon}{M} + \frac{\eta}{M}\right),\nonumber\\
   \{\theta_2,\theta_2\}_\star&=& 2\left(\frac{\epsilon}{M} + \frac{\eta}{M}\right),\nonumber\\
    \{\theta_1,\theta_2\}_\star&=& -2i \left(\frac{\epsilon}{M}+ \frac{\eta}{M}\right).
\end{eqnarray}

If we now go to the chiral coordinates

\begin{eqnarray}
  \theta &=& \theta_1+i\theta_2,\nonumber \\
  \bar{\theta} &=& \theta_1-i\theta_2,
\end{eqnarray}
the star-anticommutators are
\begin{eqnarray}
  \{\theta,\theta\}_\star&=& -8\left(\frac{\epsilon}{M} + \frac{\eta}{M}\right),\nonumber\\
   \{\bar{\theta},\bar{\theta}\}_\star&=& 0,\nonumber\\
    \{\theta,\bar{\theta}\}_\star&=&0,
\end{eqnarray}
which is the Cliffordization in half of the coordinates (in the chiral sector) as obtained in \cite{ks} and \cite{ihl}.

It might appear at first that the bosonic sector does not get deformed at all, but this is not the case. Consider the bosonic Hermitian operator

\begin{equation}
    W=\frac{i}{2}(\widehat{Q}_1\widehat{Q}_2-\widehat{Q}_2\widehat{Q}_1).
\end{equation}

We now proceed to deform it, setting $\eta=0$ for simplicity. Since $$[\overline{Q},W]=-2H\overline{Q},$$ we obtain that $W^\mathcal{F}=W$, so that $W$ undergoes no deformation. However, its coproduct exhibits nontrivial deformation:

\begin{equation}
    \Delta^\mathcal{F}(W)=\Delta(W)-\frac{2\epsilon}{M}(\overline{Q}\otimes\overline{Q}H+\overline{Q}H\otimes\overline{Q}).
\end{equation}

\subsubsection{Factorization Method}

We can also approach the question of supersymmetry deformation within the framework of the factorization method, which is an useful method for generating classes of solvable potentials for a Hamiltonian. It was devised by Infeld and Hull \cite{ih} after pioneering works by Dirac (factorization of the Hamiltonian of the harmonic oscillator) and Schr\"odinger (factorization of the radial part of the Coulomb Hamiltonian) and further generalized by Mielnik \cite{mielnik}. The method (on its first-order version) consists of factorizing the Hamiltonian by introducing intertwining operators
\begin{equation}
    A=\left(\frac{d}{dx}+\alpha(x) \right),\quad   A^+=\left(-\frac{d}{dx}+\alpha(x) \right),
\end{equation}
where $\alpha$ turns out to satisfy a Riccati differential equation (for a recent review see \cite{fernandez}).

Supersymmetry algebra can be built up in this setting  by writing $Q=\widehat{Q}_1+i\widehat{Q}_2$ and $\overline{Q}=\widehat{Q}_1-i\widehat{Q}_2$ used earlier in this section as

\begin{equation}
    Q=\left(\begin{array}{cc}0 & A^+ \\ 0 & 0 \\ \end{array}\right),\quad
        \overline{Q}=\left(\begin{array}{cc} 0 & 0 \\ A & 0 \\ \end{array}\right),
\end{equation}
The Hamiltonian is
\begin{equation}
    H=\frac{1}{2}\{Q,\overline{Q}\}=\frac{1}{2} \left(\begin{array}{cc} A^+A & 0 \\ 0 & AA^+ \\ \end{array}\right)=
    \left(\begin{array}{cc} H^+ & 0 \\ 0 & H^- \\ \end{array}\right).
\end{equation}

If we proceed to the same twist as before (setting $\eta=0$), we obtain
\begin{equation}
    Q^\mathcal{F}=\left(\begin{array}{cc}0 & A^+ \\ \frac{2\epsilon}{M}AA^+A & 0 \\\end{array}\right),\quad
        \overline{Q}^\mathcal{F}=\left(\begin{array}{cc} 0 & 0 \\ A & 0 \\ \end{array} \right),
\end{equation}
which is a non-upper triangular form for the supersymmetry generator, still satisfying
\begin{equation}
    \{Q^\mathcal{F},\overline{Q}^\mathcal{F} \}=\left(\begin{array}{cc}A^+A & 0 \\0 & AA^+ \\ \end{array}\right)=2H.
\end{equation}

\subsection{The ${\cal N}=4$ case}

We now turn to the ${\cal N}=4$ supersymmetry algebra

\begin{eqnarray}
  \{\widehat{Q}_I, \widehat{Q}_J\} &=&\delta_{IJ}H,\nonumber \\
    \left[H,\widehat{Q}_I\right]&=&0,
\end{eqnarray}
($I,J=1,2,3,4$)
and apply the twist

\begin{equation}
    \mathcal{F}=\exp\left(\frac{\eta_{ij}}{M}\overline{Q}_i\otimes\overline{Q}_j \right),
\end{equation}

where $\eta_{ij}$ is diagonal and
\begin{eqnarray}
  \overline{Q}_1 &=& \widehat{Q}_1-i\widehat{Q}_2,\nonumber \\
  \overline{Q}_2 &=& \widehat{Q}_3-i\widehat{Q}_4.
\end{eqnarray}

From now on we set $\eta_{11}=\epsilon$ and $\eta_{22}=\eta$. Being Abelian, this twist trivially satisfies the cocycle condition.

A similar procedure as before yields the deformation of the generators:

\begin{eqnarray}
  \widehat{Q}_1^\mathcal{F} &=& \widehat{Q}_1+\frac{\epsilon}{M}(\widehat{Q}_1-i\widehat{Q}_2)H,\nonumber \\
  \widehat{Q}_2^\mathcal{F} &=& \widehat{Q}_2-\frac{i\epsilon}{M}(\widehat{Q}_1-i\widehat{Q}_2)H,\nonumber \\
  \widehat{Q}_3^\mathcal{F} &=& \widehat{Q}_3+\frac{\eta}{M}(\widehat{Q}_3-i\widehat{Q}_4)H,\nonumber  \\
  \widehat{Q}_4^\mathcal{F} &=& \widehat{Q}_4-\frac{i\eta}{M}(\widehat{Q}_3-i\widehat{Q}_4)H.
\end{eqnarray}
The deformed coproducts are

\begin{eqnarray}
  \Delta^\mathcal{F}(\widehat{Q}_1) &=& \Delta(\widehat{Q}_1)+\frac{\epsilon}{M}(\widehat{Q}_1\otimes H-H\otimes \widehat{Q}_1)-\frac{i\epsilon}{M}(\widehat{Q}_2\otimes H-H\otimes \widehat{Q}_2),\nonumber \\
   \Delta^\mathcal{F}(\widehat{Q}_2) &=& \Delta(\widehat{Q}_2)-\frac{i\epsilon}{M}(\widehat{Q}_1\otimes H-H\otimes \widehat{Q}_1)-\frac{\epsilon}{M}(\widehat{Q}_2\otimes H-H\otimes \widehat{Q}_2),\nonumber\\
   \Delta^\mathcal{F}(\widehat{Q}_3) &=& \Delta(\widehat{Q}_3)+\frac{\eta}{M}(\widehat{Q}_3\otimes H-H\otimes \widehat{Q}_3)-\frac{i\eta}{M}(\widehat{Q}_4\otimes H-H\otimes \widehat{Q}_4),\nonumber\\
 \Delta^\mathcal{F}(\widehat{Q}_4) &=& \Delta(\widehat{Q}_4)-\frac{i\eta}{M}(\widehat{Q}_3\otimes H-H\otimes \widehat{Q}_3)-\frac{\eta}{M}(\widehat{Q}_4\otimes H-H\otimes \widehat{Q}_4).
\end{eqnarray}

The universal $\mathcal{R}$-matrix is simply $\mathcal{F}^{-2}$, allowing us to calculate the deformed coproducts of the deformed quantities:
\begin{eqnarray}
  \Delta^\mathcal{F}(\widehat{Q}_1^\mathcal{F}) &=& \widehat{Q}_1^\mathcal{F}\otimes\mathbf{1}+\mathbf{1}\otimes \widehat{Q}_1^\mathcal{F}+\frac{2\epsilon}{M}(\widehat{Q}_1-i\widehat{Q}_2)\otimes H,\nonumber \\
  \Delta^\mathcal{F}(\widehat{Q}_2^\mathcal{F}) &=& \widehat{Q}_2^\mathcal{F}\otimes\mathbf{1}+\mathbf{1}\otimes \widehat{Q}_2^\mathcal{F}-\frac{2i\epsilon}{M}(\widehat{Q}_1-i\widehat{Q}_2)\otimes H,\nonumber   \\
  \Delta^\mathcal{F}(\widehat{Q}_3^\mathcal{F}) &=& \widehat{Q}_3^\mathcal{F}\otimes\mathbf{1}+\mathbf{1}\otimes \widehat{Q}_3^\mathcal{F}+\frac{2\eta}{M}(\widehat{Q}_3-i\widehat{Q}_4)\otimes H,\nonumber   \\
  \Delta^\mathcal{F}(\widehat{Q}_4^\mathcal{F}) &=& \widehat{Q}_4^\mathcal{F}\otimes\mathbf{1}+\mathbf{1}\otimes \widehat{Q}_4^\mathcal{F}-\frac{2i\eta}{M}(\widehat{Q}_3-i\widehat{Q}_4)\otimes H .
\end{eqnarray}

The antipodes are
\begin{eqnarray}
  S(\widehat{Q}_1^\mathcal{F}) &=& -\widehat{Q}_1^\mathcal{F}+\frac{2\epsilon}{M}(\widehat{Q}_1-i\widehat{Q}_2),\nonumber \\
  S(\widehat{Q}_2^\mathcal{F}) &=& -\widehat{Q}_2^\mathcal{F}-\frac{2i\epsilon}{M}(\widehat{Q}_1-i\widehat{Q}_2),\nonumber  \\
  S(\widehat{Q}_3^\mathcal{F}) &=& -\widehat{Q}_3^\mathcal{F}+\frac{2\eta}{M}(\widehat{Q}_3-i\widehat{Q}_4),\nonumber  \\
  S(\widehat{Q}_4^\mathcal{F}) &=& -\widehat{Q}_4^\mathcal{F}-\frac{2i\eta}{M}(\widehat{Q}_3-i\widehat{Q}_4),
\end{eqnarray}
so that the deformed brackets are
\begin{eqnarray}
  \{\widehat{Q}_I^\mathcal{F}, \widehat{Q}_J^\mathcal{F}\}_\mathcal{F} &=&\delta_{IJ}H^\mathcal{F},\nonumber \\
    \left[H^\mathcal{F},\widehat{Q}_I^\mathcal{F}\right]_\mathcal{F}&=&0.
\end{eqnarray}

We shall now find out the action of the deformed multiplication $m^\mathcal{F}$ on a module consisting of Grassmann variables $\theta_I$, $I=1,\ldots,4$, such that  $ \{\widehat{Q}_I,\theta_J\}= \delta_{IJ}$.

Defining the star product as previously, we obtain that

\begin{eqnarray}
  \theta_1\star\theta_1&=& -\frac{\epsilon}{M},\nonumber \\
   \theta_2\star\theta_2&=& \frac{\epsilon}{M},\nonumber \\
    \theta_3\star\theta_3&=& -\frac{\eta}{M},\nonumber \\
   \theta_4\star\theta_4&=& \frac{\eta}{M};
\end{eqnarray}
all the other products coincide with the ordinary product.

If we now go to chiral coordinates

\begin{eqnarray}
  \zeta_1 &=& \theta_1+i\theta_2,\nonumber \\
  \overline{\zeta_1} &=& \theta_1-i\theta_2,\nonumber\\
  \zeta_2 &=& \theta_3+i\theta_4,\nonumber \\
  \overline{\zeta_2} &=& \theta_3-i\theta_4,
\end{eqnarray}
the star-anticommutators are
\begin{eqnarray}
  \{\zeta_I,\zeta_J\}_\star&=& -8\frac{\eta_{IJ}}{M},\nonumber\\
   \{\overline{\zeta}_I,\overline{\zeta}_J\}_\star&=& 0,\nonumber\\
    \{\zeta_I,\overline{\zeta}_J\}_\star&=&0,
\end{eqnarray}
which is again the Cliffordization of the unbarred chiral coordinates as in  \cite{ks} and \cite{ihl}.

We proceed analogously and introduce the bosonic Hermitian operators
\begin{eqnarray}
    W_1 &=& \frac{i}{2}(\widehat{Q}_1\widehat{Q}_2-\widehat{Q}_2\widehat{Q}_1) ,\nonumber\\
     W_2 &=& \frac{i}{2}(\widehat{Q}_3\widehat{Q}_4-\widehat{Q}_4\widehat{Q}_3).
\end{eqnarray}

Their algebra with the $\overline{Q}_i$'s is

\begin{equation}
    [\overline{Q}_i,W_j]=-2H\overline{Q}_i\delta_{ij} \text{    (no sum on i)},
\end{equation}
so that we obtain no deformation at the algebraic level, i.e, $W_i^\mathcal{F}=W_i$. On the other hand, their deformed coproducts read

\begin{equation}
    \Delta^\mathcal{F}(W_i)=\Delta(W_i)-\frac{2\eta_{ij}}{M}(\overline{Q}_j\otimes\overline{Q}_jH+\overline{Q}_jH\otimes\overline{Q}_j).
\end{equation}

\section{The Bosonic Heisenberg Twist}

We start with Heisenberg algebra $h_B(N)$ , with generators $x_i, p_i, \hbar$ ($i=1,2,\ldots , N$) and non-vanishing commutation relations given by
\begin{equation}
\relax [ x_i,p_j]=i\delta_{ij}\hbar.
\end{equation}

We enlarge it by introducing the elements

\begin{eqnarray}
  K_{ij} &=& \frac{p_ip_j}{\hbar}, \nonumber\\
  M_{ij} &=& \frac{x_ip_j}{\hbar}, \nonumber\\
    N_{ij} &=& \frac{p_ix_j}{\hbar}, \nonumber\\
      V_{ij} &=& \frac{x_ix_j}{\hbar},
\end{eqnarray}
which are now declared to be primitive elements of an enlarged algebra.

The thus enlarged algebra satisfies the relations

\begin{eqnarray}
  \left[K_{ij}, x_k\right] &=& -i\delta_{ik}p_j-i\delta_{jk}p_i ,\nonumber \\
  \left[M_{ij}, x_k\right] &=& -i\delta_{jk}x_i ,\nonumber\\
  \left[N_{ij},x_k\right] &=& -i\delta_{ik}x_j ,\nonumber\\
  \left[M_{ij},p_k\right] &=& i\delta_{ik}p_j ,\nonumber\\
  \left[N_{ij},p_k\right] &=& i\delta_{jk}p_i ,\nonumber\\
  \left[V_{ij},p_k\right] &=& i\delta_{ik}x_j+i\delta_{jk}x_i ,\nonumber\\
  \left[V_{ij},K_{kl}\right] &=& i\delta_{jk}M_{il}+i\delta_{jl}M_{ik}+i\delta_{ik}N_{lj}+i\delta_{il}N_{kj} ,\nonumber\\
  \left[V_{ij},M_{kl}\right] &=& i\delta_{il}V_{jk}+i\delta_{jl}V_{ik} ,\nonumber\\
  \left[V_{ij},N_{kl}\right] &=& i\delta_{ik}V_{jl}+i\delta_{jk}V_{il} ,\nonumber\\
  \left[K_{ij},M_{kl}\right] &=& -i\delta_{ik}K_{jl}-i\delta_{jk}K_{il} ,\nonumber\\
  \left[K_{ij},N_{kl}\right] &=& -i\delta_{il}K_{jk}-i\delta_{jl}K_{ik}, \nonumber\\
  \left[M_{ij},N_{kl}\right] &=& i\delta_{ik}M_{lj}-i\delta_{jl}M_{ik}.
\end{eqnarray}

Now let us consider the Hamiltonian given by

\begin{equation}
    H=\sum_i \frac{p_i^2}{2}+\omega^2\sum_i\frac{x_i^2}{2}=\lambda\left(K_{ii}+\omega^2V_{ii} \right),
\end{equation}
$\lambda$ being a suitable dimensional normalization constant.

We now twist it by applying
\begin{equation}
    \mathcal{F}=\exp(i\alpha_{ij}p_i\otimes p_j),
\end{equation}
with $\alpha_{ij}=-\alpha_{ji}$.

The deformed Hamiltonian will be
\begin{equation}
    H^\mathcal{F}=H-2\lambda\omega^2\hbar\alpha_{ij}M_{ij}+\lambda\omega^2\hbar^2\alpha_{ij}\alpha_{ik}K_{jk}.
\end{equation}

The deformed coproduct of the Hamiltonian is
\begin{equation}
    \Delta^{\cal F}(H)=\Delta(H)-2\lambda\omega^2\alpha_{ij}(p_i\otimes x_j-x_j\otimes p_i)+\lambda\omega^2\alpha_{ij}\alpha_{kj}(\hbar K_{ik}\otimes\hbar-\hbar\otimes \hbar K_{ik}),
\end{equation}
while the deformed coproduct of the deformed Hamiltonian is
\begin{equation}
\Delta^\mathcal{F}(H^\mathcal{F})=H^\mathcal{F}\otimes \mathbf{1}+\mathbf{1}\otimes
H^\mathcal{F}-4\lambda\omega^2\alpha_{ij}(p_i\otimes x_j)+2\lambda\omega^2\alpha_{ij}\alpha_{kj}(\hbar K_{ik}\otimes\hbar).
\end{equation}

\section{Conclusions}

In this work we investigated the consequences of the simplest deformation of the ${\cal N}$-extended Supersymmetric Quantum Mechanics, realized by an abelian twist of its underlying Universal Enveloping Superalgebra. We pointed out that two constructions are possible. For even values of ${\cal N}$, the $1D$ ${\cal N}$-extended superalgebra can be identified with the fermionic Heisenberg algebra. Alternatively, a realization
of the $1D$ superalgebra can be obtained in terms of operators belonging to the Universal enveloping algebra generated by one bosonic and several fermionic oscillators. We defined the deformed theories in terms of twist-deformed generators and twist-deformed brackets. We recovered, in a more general setting, the Cliffordization results of \cite{ks,ihl}.
The bosonic sector of the theory gets deformed in its multi-particle sector, even if the bosonic operators themselves get undeformed, due to their deformed coproduct.
 The difference between a ``bosonic'' versus a ``fermionic'' abelian twist has been pointed out.
\par
The models under considerations admit fermionic derivatives which do not obey the graded Leibniz rule.
{}~
\\{}~
\par {\large{\bf Acknowledgments}}{} ~\\{}~\par
Z. K. and F. T. are grateful to the S.N. Bose National Center for Basic Sciences of Kolkata for hospitality.
B. C. acknowledges a TWAS-UNESCO associateship appointment at CBPF
and CNPq for financial support. P. G. C.
acknowledges financial support from CNPq.
The work was supported by Edital Universal CNPq, Proc. 472903/2008-0 (P.G.C., Z.K., F.T).


\begin{thebibliography}{99}
\bibitem{cct}
  P.~G.~Castro, B.~Chakraborty and F.~Toppan,
  J.\ Math.\ Phys.\  {\bf 49}, 082106 (2008)
  [arXiv:0804.2936 [hep-th]].

  \bibitem{hh}
  A.~C.~Hirshfeld and P.~Henselder,
  Annals Phys.\  {\bf 302}, 59 (2002);

  A.~C.~Hirshfeld and P.~Henselder,
  Annals Phys.\  {\bf 308}, 311 (2003);

  A.~C.~Hirshfeld, P.~Henselder and T.~Spernat,
  Annals Phys.\  {\bf 314}, 75 (2004)
  [arXiv:quant-ph/0404168].

\bibitem{ks}
  Y.~Kobayashi and S.~Sasaki,
  Int.\ J.\ Mod.\ Phys.\  A {\bf 20}, 7175 (2005)
  [arXiv:hep-th/0410164].


\bibitem{ihl}
  M.~Ihl and C.~S\"amann,
  JHEP {\bf 0601}, 065 (2006)
  [arXiv:hep-th/0506057].


  \bibitem{ckt} B. Chakraborty, Z. Kuznetsova and F. Toppan, ``Twist Deformations of Rotationally Invariant Quantum Mechanics'', preprint CBPF-NF-013/09.

  \bibitem{zup}
  B.~M.~Zupnik,
  Phys.\ Lett.\  B {\bf 627}, 208 (2005)
  [arXiv:hep-th/0506043].

\bibitem{krt}
  Z.~Kuznetsova, M.~Rojas and F.~Toppan,
  Int.\ J.\ Mod.\ Phys.\  A {\bf 23}, 309 (2008)
    [arXiv:0705.4007 [hep-th]].

  \bibitem{drw}
  M.~Dimitrijevi\'c, V.~Radovanovi\'c and J.~Wess,
  JHEP {\bf 0712}, 059 (2007)
  [arXiv:0710.1746 [hep-th]].

\bibitem{svn}
  J.~H.~Schwarz and P.~Van Nieuwenhuizen,
  Lett.\ Nuovo Cim.\  {\bf 34}, 21 (1982).

  \bibitem{fl}
  S.~Ferrara and M.~A.~Lled\'o,
  JHEP {\bf 0005}, 008 (2000)
  [arXiv:hep-th/0002084].

  \bibitem{flm}
  S.~Ferrara, M.~A.~Lled\'o and O.~Macia,
  JHEP {\bf 0309}, 068 (2003)
  [arXiv:hep-th/0307039].

      \bibitem{kpt}
  D.~Klemm, S.~Penati and L.~Tamassia,
  Class.\ Quant.\ Grav.\  {\bf 20}, 2905 (2003)
  [arXiv:hep-th/0104190].

   \bibitem{bgn}
  J.~de Boer, P.~A.~Grassi and P.~van Nieuwenhuizen,
  Phys.\ Lett.\  B {\bf 574}, 98 (2003)
  [arXiv:hep-th/0302078].

\bibitem{ov1}
  H.~Ooguri and C.~Vafa,
  Adv.\ Theor.\ Math.\ Phys.\  {\bf 7}, 53 (2003)
  [arXiv:hep-th/0302109].

 \bibitem{ov2}
  H.~Ooguri and C.~Vafa,
  Adv.\ Theor.\ Math.\ Phys.\  {\bf 7}, 405 (2004)
  [arXiv:hep-th/0303063].


  \bibitem{sei}
  N.~Seiberg,
  JHEP {\bf 0306}, 010 (2003)
  [arXiv:hep-th/0305248].

\bibitem{ck}
  B.~Chandrasekhar and A.~Kumar,
  JHEP {\bf 0403}, 013 (2004)
  [arXiv:hep-th/0310137].

\bibitem{chandra}
  B.~Chandrasekhar,
  Phys.\ Rev.\  D {\bf 70}, 125003 (2004)
  [arXiv:hep-th/0408184].

\bibitem{hkks1}
  T.~Hatanaka, S.~V.~Ketov, Y.~Kobayashi and S.~Sasaki,
  Nucl.\ Phys.\  B {\bf 716}, 88 (2005)
  [arXiv:hep-th/0502026].

  \bibitem{hkks2}
  T.~Hatanaka, S.~V.~Ketov, Y.~Kobayashi and S.~Sasaki,
  Nucl.\ Phys.\  B {\bf 726}, 481 (2005)
  [arXiv:hep-th/0506071].

  \bibitem{agvm}
  L.~\'Alvarez-Gaum\'e and M.~A.~V\'azquez-Mozo,
  JHEP {\bf 0504}, 007 (2005)
  [arXiv:hep-th/0503016].

  \bibitem{fgnps}
  A.~F.~Ferrari, M.~Gomes, J.~R.~Nascimento, A.~Y.~Petrov and A.~J.~da Silva,
  Phys.\ Rev.\  D {\bf 74}, 125016 (2006)
  [arXiv:hep-th/0607087].

  \bibitem{as}
  L.~G.~Aldrovandi and F.~A.~Schaposnik,
  JHEP {\bf 0608}, 081 (2006)
  [arXiv:hep-th/0604197].

   \bibitem{is}
  E.~A.~Ivanov and A.~V.~Smilga,
  JHEP {\bf 0707}, 036 (2007)
  [arXiv:hep-th/0703038].
  
\bibitem{bal}
  E.~Akofor, A.~P.~Balachandran and A.~Joseph,
  Int.\ J.\ Mod.\ Phys.\  A {\bf 23}, 1637 (2008)
  [arXiv:0803.4351 [hep-th]].

  \bibitem{dfr}
  S.~Doplicher, K.~Fredenhagen and J.~E.~Roberts,
  Phys.\ Lett.\  B {\bf 331}, 39 (1994);

  S.~Doplicher, K.~Fredenhagen and J.~E.~Roberts,
  Commun.\ Math.\ Phys.\  {\bf 172}, 187 (1995)
  [arXiv:hep-th/0303037].

    \bibitem{sw}
  N.~Seiberg and E.~Witten,
  JHEP {\bf 9909}, 032 (1999)
  [arXiv:hep-th/9908142].

  \bibitem{cha}
  M.~Chaichian, P.~P.~Kulish, K.~Nishijima and A.~Tureanu,
  Phys.\ Lett.\  B {\bf 604}, 98 (2004)
  [arXiv:hep-th/0408069].


\bibitem{blt}
  A.~Borowiec, J.~Lukierski and V.~N.~Tolstoy,
  Mod.\ Phys.\ Lett.\  A {\bf 18}, 1157 (2003)
  [arXiv:hep-th/0301033].
  
  \bibitem{wein}
  S.~Weinberg,
  ``Quantum Field Theory'',
  Cambridge University Press, New York (1995). 

\bibitem{sor}
  E.~Sorace,
  Annales Henri Poincar\'e {\bf 3}, 659 (2002)
  [arXiv:hep-th/0205023].  
  

\bibitem{aschieri} P.~Aschieri, M.~Dimitrijevi\'c, F.~Meyer and J.~Wess,
  Class.\ Quant.\ Grav.\  {\bf 23}, 1883 (2006)
  [arXiv:hep-th/0510059].

\bibitem{bls}
  R.~Banerjee, C.~Lee and S.~Siwach,
  Eur.\ Phys.\ J.\  C {\bf 48}, 305 (2006)
  [arXiv:hep-th/0511205].
  
  \bibitem{ply}
  M.~Plyushchay,
  Int.\ J.\ Mod.\ Phys.\  A {\bf 15}, 3679 (2000)
  [arXiv:hep-th/9903130].
  
  \bibitem{ih}
  L.~Infeld and T.~E.~Hull,
  Rev.\ Mod.\ Phys.\  {\bf 23}, 21 (1951).

  \bibitem{mielnik}
  B.~Mielnik,
  J.\ Math.\ Phys.\ {\bf 25}, 3387 (1984).


  \bibitem{fernandez}
  D.~J.~Fern\'andez C.,
  arXiv:0910.0192 [quant-ph].




\end{thebibliography}
\end{document}